# Tunable high-temperature tunneling magnetoresistance in all-van der Waals antiferromagnet/semiconductor/ferromagnet junctions

*Wen Jin[1,2], Xinlu Li[4], Gaojie Zhang[1,2], Hao Wu[1,2,3], Xiaokun Wen[1,2], Li Yang[1,2], Jie Yu[1,2], Bichen Xiao[1,2], Wenfeng Zhang[1,2,3], Jia Zhang[4\*], Haixin Chang[1,2,3\*]*

[1]Center for Joining and Electronic Packaging, State Key Laboratory of Material Processing and Die & Mold Technology, School of Materials Science and Engineering, Huazhong University of Science and Technology, Wuhan 430074, China.

[2]Shenzhen R&D Center of Huazhong University of Science and Technology, Shenzhen 518000, China.

[3]Wuhan National High Magnetic Field Center and Institute for Quantum Science and Engineering, Huazhong University of Science and Technology, Wuhan 430074, China.

[4]School of Physics and Wuhan National High Magnetic Field Center, Huazhong University of Science and Technology, Wuhan 430074, China.

[\*]Corresponding author. E-mail: hxchang@hust.edu.cn, jiazhang@hust.edu.cn

**ABSTRACT**

Magnetic tunnel junctions (MTJs) have been widely applied in spintronic devices for efficient spin detection through the imbalance of spin polarization at the Fermi level. The van der Waals (vdW) nature of two-dimensional (2D) magnets with atomic-scale flat surfaces and negligible surface roughness greatly facilitates the development of

MTJs, yet is only restricted to ferromagnets. Here, we report A-type antiferromagnetism in 2D vdW single-crystal $(Fe_{0.8}Co_{0.2})_3GaTe_2$ with $T_N$~203 K in bulk and ~185 K in 9-nm nanosheets. The metallic nature and out-of-plane magnetic anisotropy make it a suitable candidate for MTJ electrodes. By constructing heterostructures based on $(Fe_{0.8}Co_{0.2})_3GaTe_2/WSe_2/Fe_3GaTe_2$, we obtain a large tunneling magnetoresistance (TMR) ratio of 180% at low temperature and the TMR retains at near-room temperature 280 K. Moreover, the TMR is tunable by the electric field down to 1 mV, implying the potential in energy-efficient spintronic devices. Our work provides new opportunities for 2D antiferromagnetic spintronics and quantum devices.

## 1. INTRODUCTION

With the rapid progress in spintronic technologies and magnetic logic, magneto-resistive random access memory (MRAM) is playing an increasingly decisive role in the information era [1-3]. As building blocks of MRAM, magnetic tunnel junctions (MTJs) enable information storage via the switch between the low and high magneto-resistance states, which correspond to parallel and antiparallel configurations of the top and bottom magnetic electrodes, and produce "0" and "1" signals [4,5]. In traditional MTJs, the magnetic electrodes usually rely on ferromagnets. Recently, with the development of antiferromagnetic (AFM) spintronics, AFM materials have drawn tremendous interest. The peculiar properties such as ultrafast terahertz (THz) spin dynamics, resistance to magnetic field disturbance, and no discernible stray field [6-8], make AFM materials suitable for next-generation devices with ultrafast speed and higher integration density [9,10].

Over the past decades, various antiferromagnets have been reported, greatly facilitating the development progress in AFM spintronics. Among them, van der Waals (vdW) AFM materials offer more opportunities in two-dimensional (2D) spintronics owing to their significant advantages of atomic-scale flat surfaces, the ability to be stacked by weak vdW interaction, and negligible surface roughness that leads to a sharp interface. However, most of them are insulating with relatively low $T_N$ (~9.5-159 K) [11-14], hindering their practical applications. To obtain metallic AFM materials, one strategy is to induce antiferromagnetism in iron-based metallic vdW ferromagnets. Investigations have been made on $Fe_nGeTe_2$ (n=3,4,5) with Co doping, and big improvements in $T_N$ were achieved by adjusting the layer stacking and chemical substitution [15-17], but simultaneously the magnetic anisotropy of the obtained AFM materials is changed to the easy plane, which is not suitable for the vertically stacked heterostructure in MTJ. Another 2D vdW metallic ferromagnet, $Fe_3GaTe_2$ (FGaT), with large perpendicular magnetic anisotropy and a high $T_C$ of about 350-380 K was reported recently [18], the weak interlayer interactions across the vdW gap provide the opportunity to adjust the interlayer coupling, which, is expected to be an ideal candidate for the parent structure to be doped.

Here, we report a vdW A-type AFM single crystals by the approach of Co doping, $(Fe_{0.8}, Co_{0.2})_3GaTe_2$ (FCGaT), with $T_N$~203 K in bulk and ~185 K in 9 nm nanosheets. By adopting FCGaT as the magnetic electrode, we realized a large tunneling magnetoresistance of 180 % at 2 K in an all-vdW antiferromagnet/semiconductor/ferromagnet junction. The TMR fades with increasing temperature but remains at 280 K. Moreover, the TMR ratio is sensitive to the electric field and its polarity can be tuned by the electric field. Our work paves the way for future AFM spintronic applications and quantum devices.

## 2. RESULTS AND DISCUSSION

**Characterizations of FCGaT**

VdW FCGaT single crystals were synthesized by a self-flux method, the details are demonstrated in the Experimental Section. Figure 1a depicts the structure of FCGaT, which is a hexagonal structure belonging to space group P63/mmc, consistent with the pristine FGaT. In FCGaT crystal, about 20% of Fe atoms are substituted by Co atoms. The top and bottom layers of Te atoms are separated by $(Fe, Co)_3Ga$ slabs, while the adjacent layers are connected by vdW force, exhibiting a vdW layered structure. The crystal structure is identified by the X-ray diffraction (XRD) analysis, as shown in Figure 1b. The sharp peaks in the XRD patterns reveal the orientation of the FCGaT crystals belonging to the (00l) crystallographic plane and indicate the high crystalline quality of the synthesized single crystals. The plate-like FCGaT single crystal photograph is illustrated in the inset of Figure 2b, demonstrating a typical dimension of

3 mm×3 mm×0.05 mm. Figure 1c shows the field emission scanning electron microscopy (FESEM) characterization results of a representative FCGaT nanoflake. From the elemental mapping, it is observed that Fe, Co, Ga, and Te atoms are distributed homogeneously in the whole nanoflake. The atomic ratio of Fe, Co, Ga, and Te is identified by energy-dispersive spectroscopy (EDS) measurements of electron probe micro-analyzer (EPMA), which is approximately 2.42: 0.62: 0.98: 1.72 (Figure S1), implying the obtained single crystal is $(Fe_{0.8}, Co_{0.2})_3GaTe_2$. The crystal has a Te deficiency of ~14%, similar to pristine FGaT [19].

**Magnetic properties of FCGaT**

To determine the magnetic properties of the FCGaT crystals, we conducted the vibrating sample magnetometer (VSM) measurement in out-of-plane ($H /\!/ c$) and in-plane ($H /\!/ ab$) magnetic field orientations, respectively. In Figure 2a, a clear peak of the zero-field-cooling (ZFC) and field-cooling (FC) curves measured under a magnetic field of 1000 Oe reveals the AFM transition, demonstrating the Neel temperature $T_N$ ~ 203 K. The sharper peak with $H /\!/ c$ indicates the magnetic moments are parallelly aligned to the c-axis. Figures 2b and d show the isothermal magnetization curves measured under the out-of-plane and in-plane magnetic fields, respectively. At T=2 K (Figures 2c), the moment of the magnetization curve shows a flatlining increase and suddenly jumps to 43 emu/g at around 4 T, then with the increasing magnetic field, the magnetization finally approaches saturation, indicative of a spin-flop transition from the AFM to FM state induced by field. Clear hysteresis loops were observed in Figure 2c, implying the existence of the FM state, though suppressed by the AFM state.

Next, we fabricated the Hall device using FCGaT nanoflakes mechanically exfoliated from the bulk single crystals to examine if the magnetism still exists in the 2D limit. The magneto-transport properties were investigated under a magnetic field perpendicular to the ab plane. As depicted in Figure 3a, the Hall resistance $R_{xy}$ as a function of the magnetic field B was measured at various temperatures. Upon sweeping the magnetic field from -9 T to 9 T, similar $R_{xy}$ versus B curves were observed with those of the bulk case (Figure 2b). Meanwhile, compared to its bulk value of ~203 K, the $T_N$ shows a slight decrease to ~185 K in nanoflakes, which is common in magnets when they are down to the 2D limit [20-23]. The temperature-dependent longitudinal resistance $R_{xx}$ is depicted in Figure 3b, as temperature decreases, the $R_{xx}$ declines monotonously, showing a typical metallic feature. At low temperatures, the $R_{xx}$ exhibits a slight increase trend, which is a consequence of Kondo scattering [24,25]. The thickness of the device was unveiled by AFM analysis, as shown in Figure 3c, demonstrating the thickness is 9 nm. The conductivity (σ) of the 9 nm nanoflake can be calculated as 714.29 S/cm at 2 K, comparable with that of the pristine $Fe_3GaTe_2$ (830.74 S/cm, 3 K) [18]. These results make FCGaT a unique vdW AFM metal that is suitable for application in spin electronics, such as MTJs.

To understand the magnetic order and electronic structure of FCGaT, we employed the first-principles calculations. First, we calculated the spin-polarized bands and density of states (DOS) of the parent FGaT as a comparison (Figure S2a, d, g). The energy difference $\Delta E$ ($\Delta E = E_{AFM} - E_{FM}$) is calculated to be approximately 14.58 meV, implying the FM coupling is more stable than the AFM state, in agreement with the ferromagnetism in FGaT as previously reported [18]. In the FCGaT crystal structure, AFM ordering shows a more stable tendency than the FM order ($\Delta E = -11.69$ meV), demonstrating the AFM coupling between the adjacent layers. However, within the whole slab of FCGaT, the FM order is estimated to be far more stable than the AFM order ($\Delta E = 375.35$ meV), which verifies its intralayer coupling is ferromagnetic. These results reveal that the ground state of bulk FCGaT corresponds to the A-type AFM configuration. From the density functional theory (DFT) calculations, FCGaT exhibits a metallic feature, similar to FGaT. Besides, in contrast to FGaT, whose band structure exhibits substantial spin polarization at the Fermi level, the band structure of FCGaT is fully spin degenerate, which further confirms the antiferromagnetism in FGaT (Figure S2 f) and is consistent with the experimental results.

**Tunneling magnetoresistance in FCGaT/WSe$_2$/FGaT vdW heterostructure**

The metallic characteristics and vdW layered structure of FCGaT provide the opportunity to be adopted as a magnetic electrode in a vertical MTJ device. The AFM/semiconductor/FM vdW heterostructure is schematized in Figure 4a, with AFM being FCGaT and FM being FGaT, which serve as the bottom and top magnetic layers, respectively. A thin WSe$_2$ layer was selected as the semiconductor interlayer, acting as the barrier layer. The Raman and photoluminescence (PL) spectra of the WSe$_2$ layer are presented in Figure S3. The Raman spectrum shows two characteristic peaks at ~252 and 260 cm$^{-1}$, which correspond to E$_{2g}^1$ and A$_{1g}$ modes, respectively [26]. Both the Raman and PL measurements signify the WSe$_2$ used in the MTJ is of high quality, which guarantees the good performance of the MTJ. To protect the heterojunction from contamination and oxidation during the following measurements, the device was encapsulated by an h-BN flake. The thickness of the heterojunction was determined by atomic force microscopy characterization (Figure 4b), implying the thicknesses of FCGaT, WSe$_2$, and FGaT are 15 nm, 3 nm, and 17 nm, respectively.

The magneto-transport measurement with an out-of-plane magnetic field was conducted by applying a bias voltage of 1 mV, as shown in Figure 4c, the blue and red curves represent the increasing and decreasing external magnetic field B. Upon changing B for a forward sweep (B increasing from -5.5 T to 5.5 T, blue curve), the magnetoresistance R of the MTJ device exhibits three states, namely I, II, and III, as labeled in the figure. To begin with, R shows a relatively small value when sweeping B from -5.5 T to -3 T, corresponding to the low resistance state I (R$_L$). With continued forward sweep (-3 T<B<1 T), a small plateau is observed, which corresponds to state II. Then R suddenly jumps to a higher plateau as B exceeds 1 T, exhibiting the high resistance state III (R$_H$). Finally, R drops back to the state I at larger B (B>5 T). To better understand these resistance states, we measured the Hall resistance R$_{xy}$ of FGaT and FCGaT with the same thickness as they are in the MTJ, the normalized $R_{xy}$ (B) curves are plotted in Figure 4f. Under a perpendicular B~6 T (which is large enough to fully align the FCGaT), the magnetization directions of FCGaT and FGaT are in a parallel state, resulting in the small resistance state I. Then a transition from FM phase to AFM phase happens in FCGaT as B>-3 T, while the magnetization direction of the FM FGaT stays unchanged, thus leads to the resistance state II. When B exceeds 1 T, the magnetization direction of FM FGaT is flipped and aligned to the field, with the FCGaT holding in the AFM state, contributing to a large R (III) since the spin electrons tunneling through the two layers are largely suppressed. With a further increase of B (B>5 T), FCGaT transforms from the AFM phase to the FM phase, and the magnetization of FCGaT is fully aligned to the field subsequently, corresponding to the state I. It is worth mentioning that the resistance state II may attributed to the incomplete transformation from the FM phase to the AFM phase of FCGaT, which can be neglected compared to the high resistance state III. Therefore, the TMR ratio can be estimated to be ~180% according to the equation $TMR = (R_H - R_L)/R_L$. Meanwhile, the device yields a resistance-area product (RA) of $3.27 \times 10^{-2}$ $\Omega cm^2$ in consideration of the junction area of ~5.9 μm$^2$, which is comparable to previous studies [27-30]. The current-voltage (*I-V*) curve of the MTJ device measured at 2 K without magnetic field is depicted in Figure 4d, where the nonlinear curve verifies the typical tunneling behavior. The |*I*|-*V* curves at selected fields (-5 T and 2 T) are shown in Figure 4e, corresponding to the $R_H$ and $R_L$, respectively, which will be discussed later.

To get a deep insight into the mechanism of the TMR effect in our MTJ device, we conducted a detailed analysis combining modeling and first-principles calculations. For conventional MTJs applying FM as both the top and bottom electrodes, the TMR effect relies on the imbalance in majority and minority spins of the two electrodes in the DOS [31,32]. By comparison, we identified that the transition between the high- and low-resistance state in the FCGaT/WSe$_2$/FGaT vdW heterojunction stems from the transition from the AFM phase to the FM phase of FCGaT. In this scenario, the tunneling current primarily relies on the relative magnetization orientations of the two magnetic electrodes, akin to that of conventional MTJs. Hence, we employ Julliere's widely accepted two-current model [33] to analyze the TMR effect in our MTJ devices. The schematic of spin-polarized electron tunneling is illustrated in Figure 5. In the electron tunneling model, the transition of FCGaT from the AFM phase to the FM phase leads to a significant change in the density distribution of the tunneling electron states at the Fermi level, which can be confirmed by the DOS structure diagram in Figure S2. Consequently, this modification leads to a change in the tunneling conductance of the device, as defined by:

$$G \propto D_L^{\uparrow}(E_F) \times D_R^{\uparrow}(E_F) + D_L^{\downarrow}(E_F) \times D_R^{\downarrow}(E_F) \qquad (1)$$

Where G represents the tunneling conductance of the device, $D_L^{\uparrow\downarrow}(E_F)$ and $D_R^{\uparrow\downarrow}(E_F)$ denote the density of states distribution of majority/minority at the Fermi levels of the left (L) and right (R) electrodes, respectively. This alteration is the key factor behind the notable TMR effect observed in the MTJ device as studied here. We integrate the DOS results from first-principles calculations into the mentioned model to predict the

TMR ratio in the device. The predicted TMR ratio for FCGaT/WSe$_2$/FGaT is comparable to that of FGaT/WSe$_2$/FGaT, which signifies the validation of our calculation.

Then we focus on the bias dependence of the TMR in the MTJ device. By applying different bias currents from 1 mV to 0.8 V, we obtained a series of TMR versus B curves at 2 K, as presented in Figure 6a, and the extracted TMR ratio as a function of bias voltage is depicted in Figure 6c. It can be confirmed that the MTJ device can work stably at a various bias range, even down to 1 mV, which indicates its low-energy consumption and has prospective application in next-generation energy-efficient spintronic devices.. We note that the TMR-V curve is asymmetric concerning zero bias, which can be attributed to the asymmetric interface of the MTJ. To begin with, the TMR ratio decreases monotonically as the bias increases. With further increasing of the bias, a negative TMR ratio of 3.8% at 0.6 V is witnessed and its polarity remains negative as V>0.6 V. The polarity change of TMR at a larger bias has also been observed in FGT-based FM MTJs [34], which is caused by the localized spin states with large energy. The |$I$|-$V$ curves measured at selective magnetic fields in Figure 4 e make it more intuitive to understand the TMR polarity switching in our MTJ device driven by the bias. As shown in Figure 4e, a crossover was observed at a bias voltage of ~0.56 V, that is, below such a bias, the resistance at 2 T is higher than that at -5 T, yielding a positive TMR; when the bias keeps increasing, the resistance at -5 T is beyond that at 2 T, driving the TMR polarity to negative, which, is corresponding to the sign change plotted in Figure

6c. Similar phenomenon was also observed in another MTJ device with a thicker $WSe_2$ barrier of 6.8 nm (device B), signifying its universality in our AFM/semiconductor/FM heterojunctions.

We further characterized the magneto-transport properties at various temperatures to investigate the temperature dependence of the device, the results are presented in Figure 6b. The TMR ratios are extracted and plotted in Figure 6d as a function of temperature. As the temperature increases, the TMR ratio shows a declining trend but remains at 0.3% under near-room temperature 280 K. The existing TMR effect until 280 K indicates the transition from a FM state to a paramagnetic (PM) state in FCGaT. The decrease in the TMR ratio with increasing temperature can be attributed to the reduction of spin polarization influenced by the intensified thermal fluctuation.

## 3. CONCLUSIONS

In summary, we report a vdW A-type AFM single crystal FCGaT, the Neel temperature of which is ~203 K in bulk and slightly declines to ~185 K in 9 nm nanoflakes. The metallic behavior of FCGaT enables it to be a magnetic electrode in the MTJ device. By fabricating an all-vdW MTJ device based on FCGaT/$WSe_2$/FGaT heterojunctions, we observed a large TMR ratio of 180% at low temperature. The TMR ratio shows strong bias-dependent properties and can be well-tuned by bias voltage. Our work sheds light on the metallic vdW antiferromagnets for AFM spintronics, next-generation spintronic devices, and AFM quantum devices.

## 4. EXPERIMENTAL SECTION

**Single crystal growth**

Single crystals of FCGaT were grown by a self-flux method. Fe, Co powders, Ga lumps, and Te powders with high purity in a molar ratio of 1.7: 0.3: 1: 2 were mixed in an evacuated quartz tube and sealed. The quartz tube was fast heated to 1000 °C and kept for 1 day, then the mixture was slowly cooled down to 780 °C at a rate of 1.8 °C/h, before a natural cooling process.

**Device fabrication**

**FCGaT Hall device:** The Hall pattern with six electrodes was prepared by a laser direct writing machine (Micro Writer ML3, DMO) on Si substrates with 300 nm thick $SiO_2$. Cr/Au of 5/15 nm was deposited by an electron beam evaporation system, followed by a lift-off process. FCGaT nanoflakes were mechanically exfoliated from bulk single crystals and transferred onto the Hall electrodes via a dry-transfer method. The transfer process was conducted in a glove box with water and oxygen less than 0.1 ppm.

**FCGaT/$WSe_2$/FGaT MTJs:** FCGaT and FGaT single crystals were synthesized by a self-flux method, as mentioned above. $WSe_2$ single crystals were grown by CVT. The four-terminal electrodes were fabricated using the same process as the Hall bar pattern mentioned above. In the glove box, FCGaT, $WSe_2$, and FGaT were successively exfoliated and transferred onto the prepared electrodes to form a heterojunction.

**Characterizations**

The high crystallinity of the single crystals was identified via X-ray diffraction (XRD, Bruker D8 Advance) analysis. The doping level was determined by energy-dispersive X-ray spectroscopy (EDS). The morphology and elemental mapping were characterized using field emission scanning electron microscopy (FESEM, Gemini SEM 300). The thickness of the FCGaT, FGaT, and $WSe_2$ flakes was characterized by atomic force microscopy (AFM, XE7, Park; SPM9700, Shimadzu; Dimension EDGE, Bruker).

**Magnetization measurement**

Magnetization was measured under a magnetic field along the c-axis or ab-plane using a vibrating sample magnetometer option of the Physical Property Measurement System (PPMS, DynaCool Quantum Design). The electrical transport and magneto-transport properties were measured in a physical property measurement system (PPMS, DynaCool, Quantum Design). The magnetic field was applied perpendicular to the device.

**Theoretical calculation**

In the current study, the first-principles calculations have been performed by using the Vienna ab initio simulation package (VASP) [35]. The Projector Augmented Wave (PAW) potential [36] is applied to the elements, and the local density approximation (LDA) [37,38] exchange-correlation function has been employed as it is more suitable than the generalized gradient approximation (GGA) [39] for describing the magnetic properties of FGaT-like material [40]. Then we first investigated the electrical and magnetic properties of FCGaT, the cut-off energy of plane wave basis was set as 500 eV and the convergence of the total energy was set to be less than 10-7 eV. All atomic positions and lattice constants of FCGaT had been fully relaxed until the force on each atom is less than 0.01 eV/Å. The Brillouin zone sampling was performed by using the gamma-centered k-meshes 15 × 15 × 3, and 30 × 30 × 6 for the self-consistent calculation and density of states (DOS) calculation, respectively.

## Notes

The authors declare no competing financial interest.

## ACKNOWLEDGMENTS

This work was supported by the National Key Research and Development Program of China (No. 2022YFE0134600) and the National Natural Science Foundation of China (No. 52272152, 61674063 and 62074061), the Foundation of Shenzhen Science and Technology Innovation Committee (JCYJ20210324142010030 and JCYJ20180504170444967), Natural Science Foundation of Hubei Province, China (Grant No. 2022CFA031) and the fellowship of China Postdoctoral Science Foundation (No. 2022M711234). We thank the EDS and Raman tests from the Analytical Center of Huazhong University of Science and Technology, and XRD, SEM and AFM tests from Large-scale scientific instrument sharing platform in school of optical and electronic information of Huazhong University of Science and Technology.

# REFERENCES

[1] Wolf, S. A.; Awschalom, D. D.; Buhrman, R. A.; Daughton, J. M.; von Molnár, S.; Roukes, M. L.; Chtchelkanova, A. Y.; Treger, D. M. Spintronics: a spin-based electronics vision for the future. *Science* **2001**, *294* (5546), 1488-1495.

[3] Yang, H.; Valenzuela, S. O.; Chshiev, M.; Couet, S.; Dieny, B.; Dlubak, B.; Fert, A.; Garello, K.; Jamet, M.; Jeong, D.-E.; et al. Two-dimensional materials prospects for non-volatile spintronic memories. *Nature* **2022**, *606* (7915), 663-673.

[3] Tehrani, S.; Engel, B.; Slaughter, J. M.; Chen, E.; DeHerrera, M.; Durlam, M.; Naji, P.; Whig, R.; Janesky, J.; Calder, J. Recent developments in magnetic tunnel junction MRAM. *IEEE Transactions on Magnetics* **2000**, *36* (5), 2752-2757.

[4] Yuasa, S.; Nagahama, T.; Fukushima, A.; Suzuki, Y.; Ando, K. Giant room-temperature magnetoresistance in single-crystal Fe/MgO/Fe magnetic tunnel junctions. *Nature Materials* **2004**, *3* (12), 868-871.

[5] Parkin, S. S. P.; Kaiser, C.; Panchula, A.; Rice, P. M.; Hughes, B.; Samant, M.; Yang, S.-H. Giant tunnelling magnetoresistance at room temperature with MgO (100) tunnel barriers. *Nature Materials* **2004**, *3* (12), 862-867.

[6] Jungwirth, T.; Marti, X.; Wadley, P.; Wunderlich, J. Antiferromagnetic spintronics. *Nature Nanotechnology* **2016**, *11* (3), 231-241.

[7] Železný, J.; Wadley, P.; Olejník, K.; Hoffmann, A.; Ohno, H. Spin transport and spin torque in antiferromagnetic devices. *Nature Physics* **2018**, *14* (3), 220-228.

[8] Baltz, V.; Manchon, A.; Tsoi, M.; Moriyama, T.; Ono, T.; Tserkovnyak, Y. Antiferromagnetic spintronics. *Reviews of Modern Physics* **2018**, *90* (1), 015005.

[9] Olejník, K.; Seifert, T.; Kašpar, Z.; Novák, V.; Wadley, P.; Campion, R. P.; Baumgartner, M.; Gambardella, P.; Němec, P.; Wunderlich, J.; et al. Terahertz electrical writing speed in an antiferromagnetic memory. *Science Advances* **2018**, *4* (3), eaar3566.

[10] Liu, Z.; Feng, Z.; Yan, H.; Wang, X.; Zhou, X.; Qin, P.; Guo, H.; Yu, R.; Jiang, C. Antiferromagnetic Piezospintronics. *Advanced Electronic Materials* **2019**, *5* (7), 1900176.

[11] McGuire, M. A.; Garlea, V. O.; Kc, S.; Cooper, V. R.; Yan, J.; Cao, H.; Sales, B. C. Antiferromagnetism in the van der Waals layered spin-lozenge semiconductor CrTe3. *Physical Review B* **2017**, *95* (14), 144421.

[12] Jang, S. W.; Jeong, M. Y.; Yoon, H.; Ryee, S.; Han, M. J. Microscopic understanding of magnetic interactions in bilayer CrI3. *Physical Review Materials* **2019**, *3* (3), 031001.

[13] Chen, W.; Sun, Z.; Wang, Z.; Gu, L.; Xu, X.; Wu, S.; Gao, C. Direct observation of van der Waals stacking-dependent interlayer magnetism. *Science* **2019**, *366* (6468), 983-987.

[14] Otrokov, M. M.; Klimovskikh, I. I.; Bentmann, H.; Estyunin, D.; Zeugner, A.; Aliev, Z. S.; Gaß, S.; Wolter, A. U. B.; Koroleva, A. V.; Shikin, A. M.; et al. Prediction and observation of an antiferromagnetic topological insulator. *Nature* **2019**, *576* (7787), 416-422.

[15] May, A. F.; Du, M.-H.; Cooper, V. R.; McGuire, M. A. Tuning magnetic order in the van der Waals metal $Fe_5GeTe_2$ by cobalt substitution. *Physical Review Materials* **2020**, *4* (7), 074008.

[16] Seo, J.; An, E. S.; Park, T.; Hwang, S.-Y.; Kim, G.-Y.; Song, K.; Noh, W.-s.; Kim, J. Y.; Choi, G. S.; Choi, M.; et al. Tunable high-temperature itinerant antiferromagnetism in a van der Waals magnet. *Nature Communications* **2021**, *12* (1), 2844.

[17] Zhang, H.; Raftrey, D.; Chan, Y.-T.; Shao, Y.-T.; Chen, R.; Chen, X.; Huang, X.; Reichanadter, J. T.; Dong, K.; Susarla, S.; et al. Room-temperature skyrmion lattice in a layered magnet $(Fe_{0.5}Co_{0.5})_5GeTe_2$. *Science Advances 8* (12), eabm7103.

[18] Zhang, G.; Guo, F.; Wu, H.; Wen, X.; Yang, L.; Jin, W.; Zhang, W.; Chang, H. Above-room-temperature strong intrinsic ferromagnetism in 2D van der Waals $Fe_3GaTe_2$ with large perpendicular magnetic anisotropy. *Nature Communications* **2022**, *13* (1), 5067.

[19] Zhang, G.; Luo, Q.; Wen, X.; Wu, H.; Yang, L.; Jin, W.; Li, L.; Zhang, J.; Zhang, W.; Shu, H.; et al. Giant 2D Skyrmion Topological Hall Effect with Ultrawide Temperature Window and Low-Current Manipulation in 2D Room-Temperature Ferromagnetic Crystals. *Chinese Physics Letters* **2023**, *40* (11), 117501.

[20] Burch, K. S.; Mandrus, D.; Park, J.-G. Magnetism in two-dimensional van der Waals materials. *Nature* **2018**, *563* (7729), 47-52.

[21] Deiseroth, H.-J.; Aleksandrov, K.; Reiner, C.; Kienle, L.; Kremer, R. K. $Fe_3GeTe_2$ and $Ni_3GeTe_2$-Two New Layered Transition-Metal Compounds: Crystal

Structures, HRTEM Investigations, and Magnetic and Electrical Properties. *European Journal of Inorganic Chemistry* **2006**, *2006* (8), 1561-1567.

[22] Lee, J.-U.; Lee, S.; Ryoo, J. H.; Kang, S.; Kim, T. Y.; Kim, P.; Park, C.-H.; Park, J.-G.; Cheong, H. Ising-Type Magnetic Ordering in Atomically Thin FePS$_3$. *Nano Letters* **2016**, *16* (12), 7433-7438.

[23] Tian, Y.; Gray, M. J.; Ji, H.; Cava, R. J.; Burch, K. S. Magneto-elastic coupling in a potential ferromagnetic 2D atomic crystal. *2D Materials* **2016**, *3* (2), 025035.

[24] Choi, D.-J.; Lorente, N. Magnetic Impurities on Surfaces: Kondo and Inelastic Scattering. In *Handbook of Materials Modeling: Applications: Current and Emerging Materials*, Andreoni, W., Yip, S. Eds.; Springer International Publishing, 2020; pp 467-498.

[25] Martino, E.; Putzke, C.; König, M.; Moll, P. J. W.; Berger, H.; LeBoeuf, D.; Leroux, M.; Proust, C.; Akrap, A.; Kirmse, H.; et al. Unidirectional Kondo scattering in layered NbS$_2$. *npj 2D Materials and Applications* **2021**, *5* (1), 86.

[26] Zeng, H.; Liu, G.-B.; Dai, J.; Yan, Y.; Zhu, B.; He, R.; Xie, L.; Xu, S.; Chen, X.; Yao, W.; et al. Optical signature of symmetry variations and spin-valley coupling in atomically thin tungsten dichalcogenides. *Scientific Reports* **2013**, *3* (1), 1608.

[27] Jin, W.; Zhang, G.; Wu, H.; Yang, L.; Zhang, W.; Chang, H. Room-temperature spin-valve devices based on Fe$_3$GaTe$_2$/MoS$_2$/Fe$_3$GaTe$_2$ 2D van der Waals heterojunctions. *Nanoscale* **2023**, 15 (11), 5371-5378,

[28] Yin, H.; Zhang, P.; Jin, W.; Di, B.; Wu, H.; Zhang, G.; Zhang, W.; Chang, H. Fe$_3$GaTe$_2$/MoSe$_2$ ferromagnet/semiconductor 2D van der Waals heterojunction

for room-temperature spin-valve devices. *Cryst. Eng. Comm.* **2023**, 25 (9), 1339-1346

[29] Jin, W.; Zhang, G.; Wu, H.; Yang, L.; Zhang, W.; Chang, H. Room-Temperature and Tunable Tunneling Magnetoresistance in $Fe_3GaTe_2$-Based 2D van der Waals Heterojunctions. *ACS Applied Materials & Interfaces* **2023**, 15 (30), 36519-36526.

[30] Zhu, W.; Xie, S.; Lin, H.; Zhang, G.; Wu, H.; Hu, T.; Wang, Z.; Zhang, X.; Xu, J.; Wang, Y.; et al. Large Room-Temperature Magnetoresistance in van der Waals Ferromagnet/Semiconductor Junctions. *Chinese Physics Letters* **2022**, *39* (12), 128501.

[31] Bowen, M.; Cros, V.; Petroff, F.; Fert, A.; Martínez Boubeta, C.; Costa-Krämer, J. L.; Anguita, J. V.; Cebollada, A.; Briones, F.; de Teresa, J. M.; et al. Large magnetoresistance in Fe/MgO/FeCo (001) epitaxial tunnel junctions on GaAs (001). *Applied Physics Letters* **2001**, *79* (11), 1655-1657.

[32] Wulfhekel, W.; Klaua, M.; Ullmann, D.; Zavaliche, F.; Kirschner, J.; Urban, R.; Monchesky, T.; Heinrich, B. Single-crystal magnetotunnel junctions. *Applied Physics Letters* **2001**, *78* (4), 509-511.

[33] Julliere, M. Tunneling between ferromagnetic films. *Physics Letters A* **1975**, *54* (3), 225-226.

[34] Min, K.-H.; Lee, D. H.; Choi, S.-J.; Lee, I.-H.; Seo, J.; Kim, D. W.; Ko, K.-T.; Watanabe, K.; Taniguchi, T.; Ha, D. H.; et al. Tunable spin injection and detection across a van der Waals interface. *Nature Materials* **2022**, *21* (10), 1144-1149.

[35] Kresse, G.; Furthmüller, J. Efficiency of ab-initio total energy calculations for metals and semiconductors using a plane-wave basis set. *Computational Materials Science* **1996**, *6* (1), 15-50.

[36] Kresse, G.; Joubert, D. From ultrasoft pseudopotentials to the projector augmented-wave method. *Physical Review B* **1999**, *59* (3), 1758-1775.

[37] Perdew, J. P.; Zunger, A. Self-interaction correction to density-functional approximations for many-electron systems. *Physical Review B* **1981**, *23* (10), 5048-5079.

[38] Ceperley, D. M.; Alder, B. J. Ground State of the Electron Gas by a Stochastic Method. *Physical Review Letters* **1980**, *45* (7), 566-569.

[39]. Perdew, J. P.; Burke, K.; Ernzerhof, M. Generalized Gradient Approximation Made Simple. *Physical Review Letters* **1996**, *77* (18), 3865-3868.

[40]. Li, X.; Zhu, M.; Wang, Y.; Zheng, F.; Dong, J.; Zhou, Y.; You, L.; Zhang, J. Tremendous tunneling magnetoresistance effects based on van der Waals room-temperature ferromagnet Fe3GaTe2 with highly spin-polarized Fermi surfaces. *Applied Physics Letters* **2023**, *122* (8), 082404.

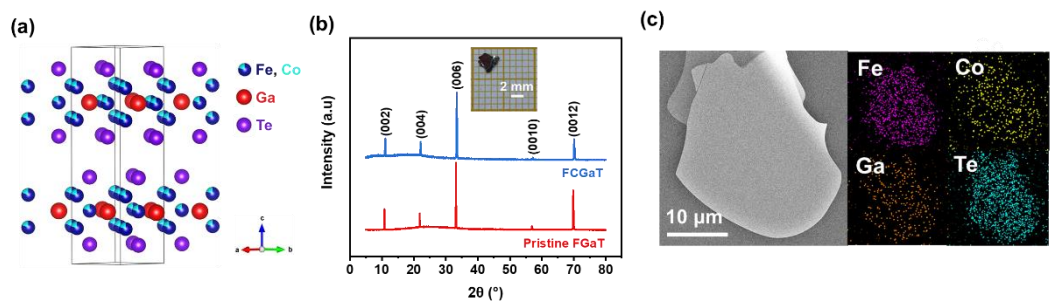

**Figure 1. Crystal structure and characterizations of FCGaT single crystals.** (a) Crystal structure of FCGaT. (b) XRD patterns of the pristine FGaT and FCGaT single crystal. Inset: a typical single-crystal photograph. (c) SEM morphology and the corresponding elemental mapping of a representative FCGaT nanoflake.

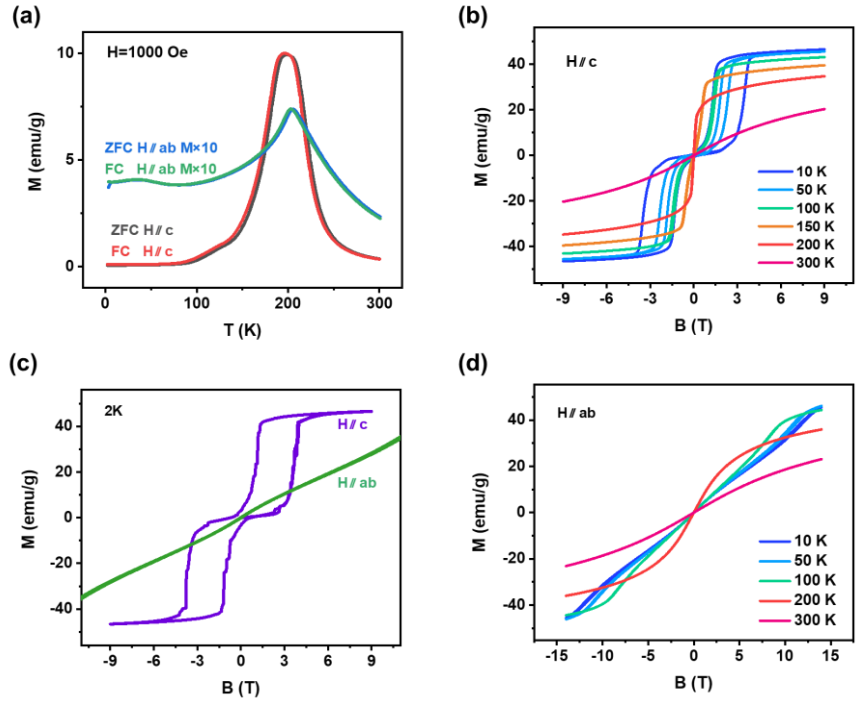

**Figure 2. Anisotropic magnetization measurements for bulk FCGaT single crystals.** (a) Temperature-dependent magnetization for the fixed applied field (B=1000 Oe, out-of-plane and in-plane). (b) Isothermal M-H curves of FCGaT bulk crystals at varying temperatures with magnetic fields along the c axis (b) and ab plane (d). (c) magnetization hysteresis loops measured at 2 K with magnetic fields along the c-axis and ab plane, respectively.

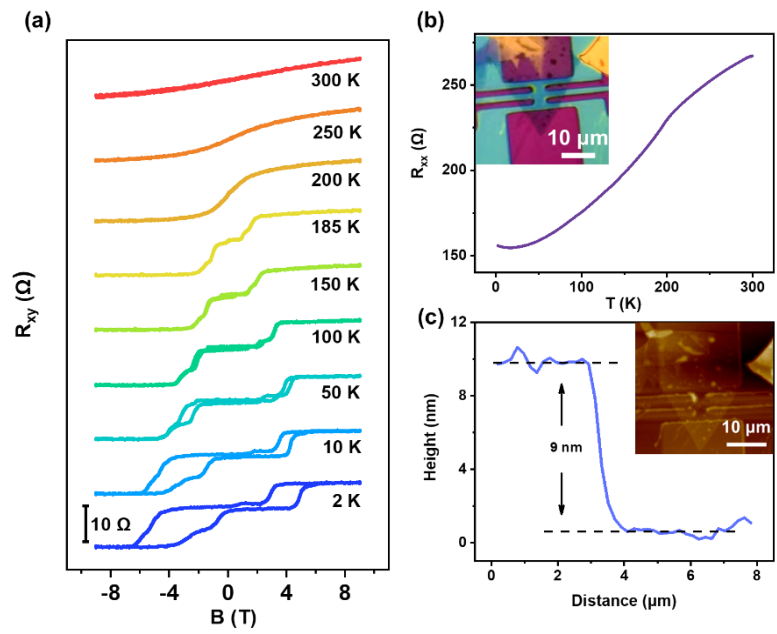

**Figure 3. Spin-flop transition in few-layer FCGaT.** (a) $R_{xy}$ at different temperatures from 2 to 300 K. (b) Temperature dependence of the $R_{xx}$. Inset: the optical image of the Hall device. (c) The AFM image and height profile of the FCGaT Hall device, showing the thickness is 9 nm.

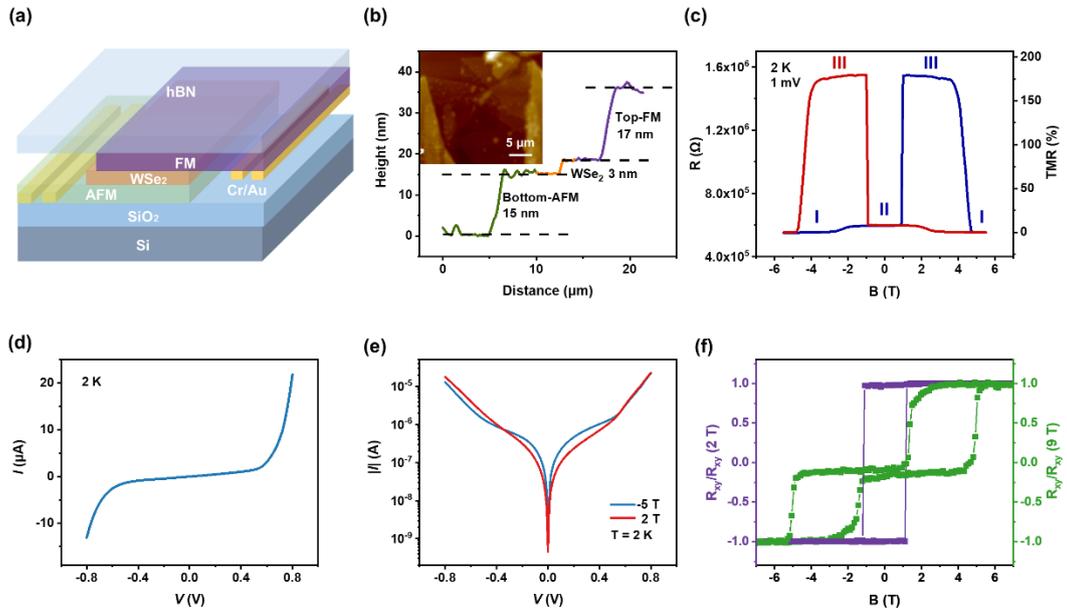

**Figure 4. Characterization of the FCGaT/WSe₂/FGaT MTJ.** (a) Schematic diagram of the FCGaT/WSe₂/FGaT MTJ device. (b) AFM height profile of a typical device, showing the thickness of the bottom FCGaT, WSe₂, and FGaT are 15, 3, and 17 nm, respectively. Inset: the AFM image of the MTJ. (c) Resistance and TMR versus perpendicular magnetic field at 2 K with a bias of 1 mV. (d) $I-V$ characteristics at 2 K. (e) $I$-$V$ curves of the device at selected magnetic fields. (f) Normalized $R_{xy}(B)$ curves at 2 K for FGaT (purple lines) and FCGaT (green lines) with similar thicknesses of the MTJ device.

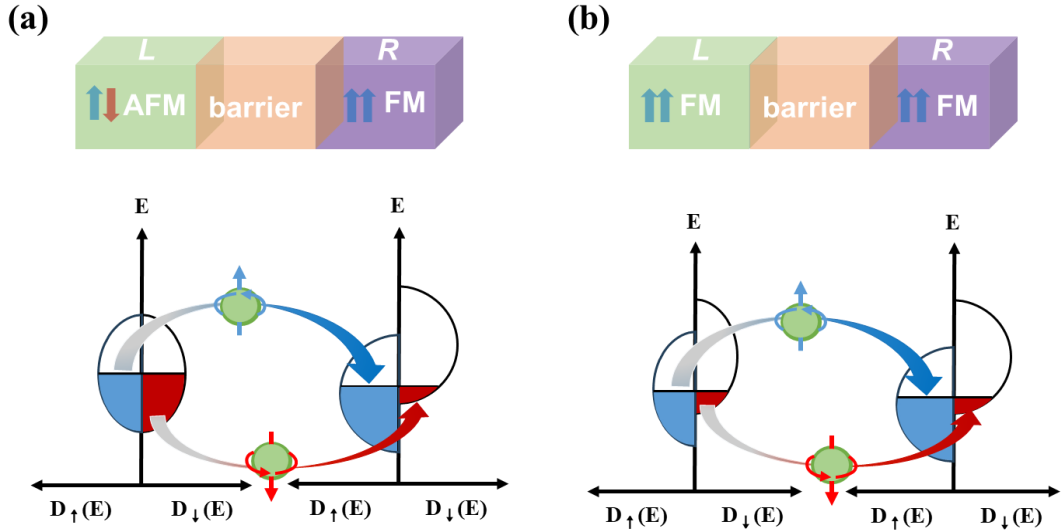

**Figure 5. Schematic diagram of spin-polarized electron tunneling for FCGaT /WSe₂/FGaT MTJs.** (a) Schematic diagrams of the electron transport in FCGaT /WSe₂/FGaT MTJs when FCGaT is in the AFM phase. (b) Schematic diagrams of the electron transport in FCGaT /WSe₂/FGaT MTJs when FCGaT is in the FM phase at large magnetic fields.

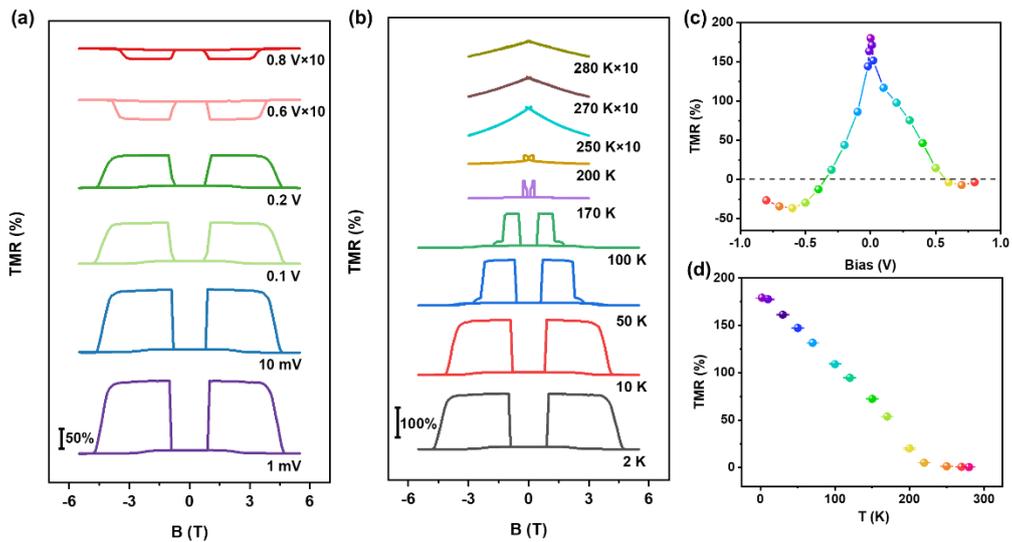

**Figure 6. Magneto-transport measurements at various bias voltages and temperatures.** (a) TMR curves measured at various bias voltages ranging from 1 mV

to 0.8 V at 2 K. (b) TMR curves measured at various temperatures ranging from 2 to 280 K with a fixed bias of 1 mV. (c) $V$-dependent TMR ratio variations for the MTJ. (d) Temperature-dependent TMR ratios extracted from panel (b).

# Supplementary Information

# Tunable high-temperature tunneling magnetoresistance in all-van der Waals antiferromagnet/semiconductor/ferromagnet junctions

Wen Jin[1,2], Xinlu Li[4], Gaojie Zhang[1,2], Hao Wu[1,2,3], Xiaokun Wen[1,2], Li Yang[1,2], Jie Yu[1,2], Bichen Xiao[1,2], Wenfeng Zhang[1,2,3], Jia Zhang[4*], Haixin Chang[1,2,3*]

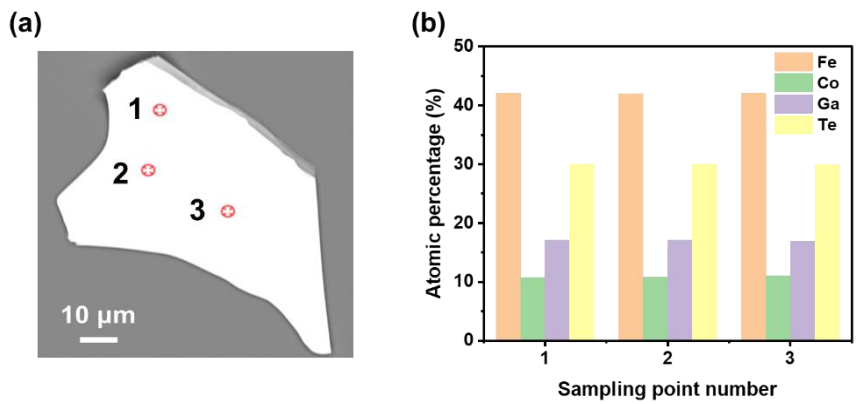

**Figure S1. EDS analysis of elemental content in a representative FCGaT nanoflake.** (a) EDS image of the tested FCGaT nanoflake with sampling points labeled 1,2 and 3, respectively. (b) The atomic ratio at % of Fe, Co, Ga, and Te for sampling point 1 (Fe:Co:Ga:Te=2.43:0.64:0.98:1.73), sampling point 2 (Fe:Co:Ga:Te=2.41:0.62:0.98:1.72), and sampling point 3 (Fe:Co:Ga:Te=2.41:0.61:0.98:1.72). The average atomic ratio at % of Fe:Co:Ga:Te is 2.42:0.62:0.98:1.72, implying a Te deficiency of ~14% of FCGaT.

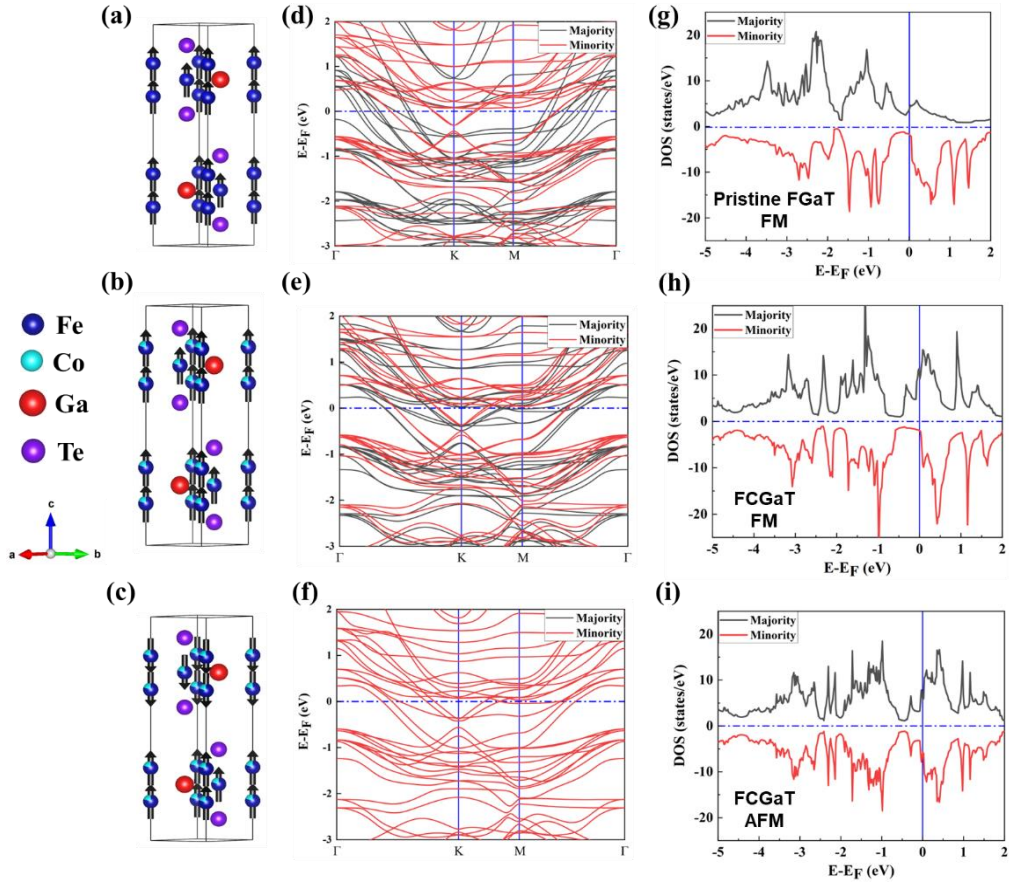

**Figure S2. Crystal structures, spin-polarized bands, and electronic density of states (DOS) structures from first-principles calculations.** (a-c) Crystal structure, (d-f) spin-polarized bands, and (g-i) DOS for the FM order in FGaT (a,d,g) and FCGaT (b,e,h), as a comparison with that for the AFM order in FCGaT (c,f,i), where AFM interaction comes from the interlayer.

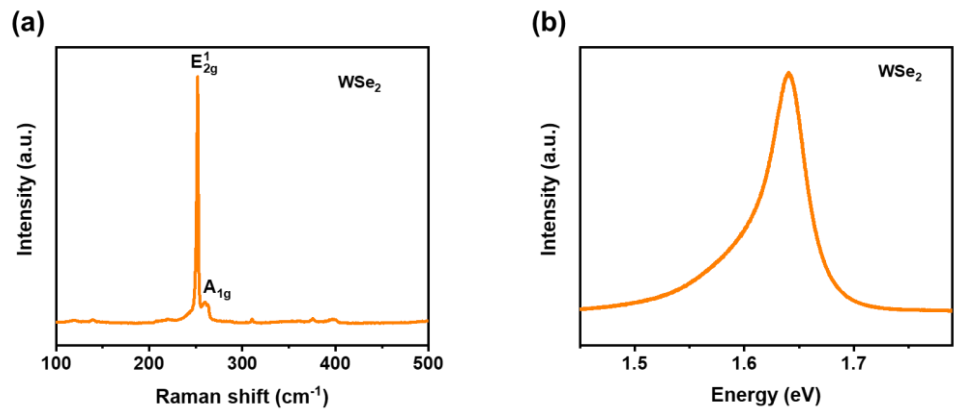

**Figure S3.** Raman and PL spectra of $WSe_2$ in the MTJ mentioned in the main text.

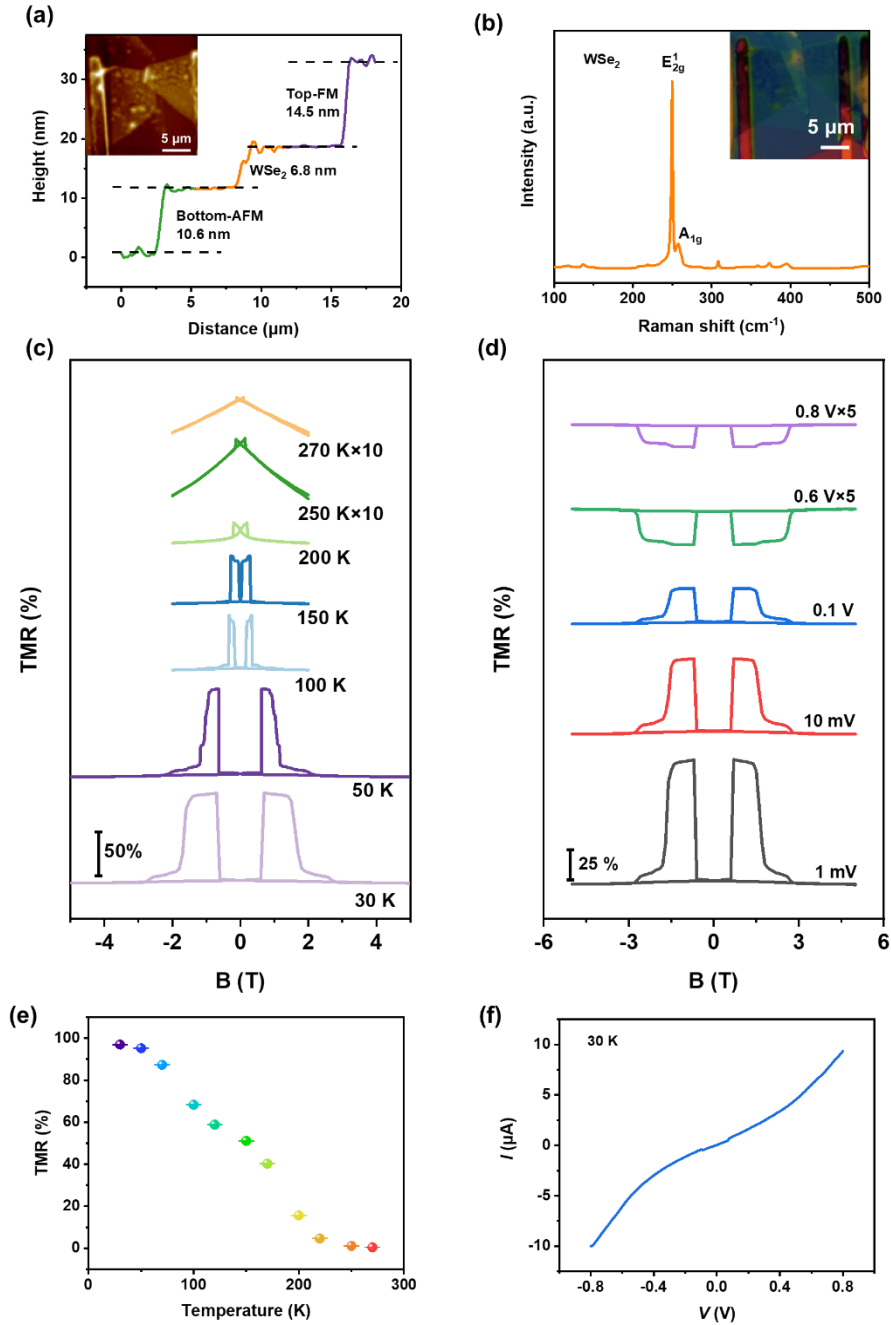

**Figure S4. Characterization, magneto-transport, and electrical properties of another MTJ device (device B).** (a) AFM height profile of device B, showing the thickness of the bottom FCGaT, WSe$_2$, and FGaT are 10.6, 6.8, and 14.5 nm, respectively. Inset: the AFM image of the device. (b) Raman spectrum of WSe$_2$ (device B). (c) TMR curves measured at various temperatures ranging from 2 to 280 K with a fixed bias of 1 mV. (d) TMR curves measured at various bias voltages ranging from 1

mV to 0.8 V at 30 K. (e) Temperature-dependent TMR ratios extracted from panel (c). (f) $I-V$ characteristics at 30 K.